\documentclass[a4paper]{jpconf}
\usepackage{graphicx}
\usepackage{amssymb}
\usepackage{cite}
\usepackage{caption}

\begin{document}
\title{Storage-ring ionization and recombination experiments with multiply charged ions relevant to astrophysical and fusion plasmas}

 \author{Stefan Schippers}

 \address{Institut f\"{u}r Atom- und Molek\"{u}lphysik, Justus-Liebig-Universit\"{a}t, 35392 Giessen, Germany}

 \ead{Stefan.Schippers@iamp.physik.uni-giessen.de}

\begin{abstract}
Past and ongoing research activities at the Heidelberg heavy-ion storage-ring TSR are reviewed which aim at providing accurate absolute rate coefficients and cross sections of atomic collision processes for applications in astrophysics and magnetically confined fusion. In particular, dielectronic recombination and electron impact ionization of iron ions are discussed as well as dielectronic recombination of tungsten ions.
\end{abstract}

\section{Introduction}

Electron-impact ionization (EII), photoionization (PI), and dielectronic recombination (DR) are important atomic collision processes since they are governing the charge balance in atomic plasmas. Accurate cross sections and rate coefficients are therefore required for these processes --- along with many other atomic data --- for the interpretation of observations of such plasmas be they man-made or astrophysical. Because of the vast atomic data needs most of the data that are presently used in plasma modeling codes have been generated by theoretical calculations. In order to assess the reliability of these calculations and to point out directions for their improvements benchmarking experiments are vitally needed \cite{Ferland2003a,Kallman2007a}.

For more than a decade, measurements of absolute DR rate coefficients have been performed employing the electron-ion merged-beams method at the heavy-ion storage ring TSR of the Max-Planck-Institute for Nuclear Physics in Heidelberg, Germany. These investigations focussed on iron ions because of their prominent role in X-ray astronomy. Iron is the most abundant heavy element and still contributes to line emission from astrophysical plasmas when lighter elements are already fully stripped. The present status of the experimentally-derived Fe-DR data base has been summarized recently \cite{Schippers2010}. A more comprehensive tabulation of storage-ring recombination experiments with astrophysically relevant ions can be found in \cite{Schippers2009a}.

Here, the experimental procedure for measuring \emph{absolute} rate coefficients and cross sections for DR and EII at a heavy-ion storage ring is briefly outlined. Then, selected results for EII of iron ions and DR of iron and tungsten ions are presented and discussed in the contexts of astrophysical and fusion plasmas. Throughout the paper, ions are identified by their primary charge state before EII or DR.

\section{Experimental method}

The experimental procedures for measuring recombination rate coefficients and electron-impact ionization cross-sections at the TSR heavy-ion storage ring \cite{Wolf2006c} have been described comprehensively before (see e.g.\ \cite{Mueller1999c,Schippers2001c,Schmidt2007b,Lestinsky2009,Hahn2010}, and references therein). Briefly, ions with well defined mass and charge state from an accelerator are injected into the storage ring. In one of the straight sections of the storage ring an electron beam is magnetically guided onto the ion beam such that both beams are centered on each other and move collinearly over a distance $L \approx 1.5$~m. Electron-ion collisions occurring in this merged-beams overlap region may lead to ionization or recombination of the primary ions. The product ions are separated from the primary beam in the first storage-ring dipole magnet behind the electron target and are counted by suitably positioned single particle detectors with efficiency $\eta$ of practically 100\%.

Absolute merged-beams rate coefficients for recombination or ionization are readily derived by normalizing the detected count rate $R$ to the stored ion current $I_i$  and to the electron density $n_e$, i.e.\
\begin{equation}\label{eq:alphaMBexp}
    \alpha_\mathrm{MB}(E_\mathrm{rel}) = \langle \sigma v_\mathrm{rel}\rangle = R\frac{eqv_i}{(1-\beta_i\beta_e)I_in_eL\eta}.
\end{equation}
Here $eq$ is the charge of the primary ion, $v_i = c\beta_i$ and $v_e= c \beta_e$ are the ion and electron velocities in the laboratory system, respectively, and $c$ denotes the speed of light in vacuum. The relative velocity $v_\mathrm{rel}$ and the corresponding  electron-ion collision energy $E_\mathrm{rel}$ are readily calculated  from $v_i$ and $v_e$ \cite{Schippers2000b}. In a TSR electron-ion recombination or ionization experiment $v_i$ is kept fixed and $E_\mathrm{rel}$ is varied by changing $v_e$ via the cathode voltage at the electron gun. At energies $E_\mathrm{rel}$ much larger than the experimental energy spread, which scales with $v_\mathrm{rel}$ and can be as low as 1~meV \cite{Wolf2006c}, the merged-beams rate coefficient can safely be converted into the apparent cross section $\sigma(E_\mathrm{rel}) =  \alpha_\mathrm{MB}(E_\mathrm{rel})/v_\mathrm{rel}$. The quantity which is most relevant for plasma physical applications is the plasma rate coefficient $\alpha(T_e)$ as function of plasma electron temperature $T_e$. It is obtained from the apparent cross section by a convolution with an isotropic Maxwellian electron energy distribution as \cite{Mueller1999c,Schippers2001c}
\begin{equation}
\alpha(T_e) = \frac{1}{(k_{B}T)^{3/2}}\sqrt{\frac{8}{m_{e}\pi}}\int_0^\infty dE_\mathrm{rel} \, \sigma(E_\mathrm{rel}) E_\mathrm{rel} \exp{(-E_\mathrm{rel}/k_{B}T_e)}
\label{eq:alphaTsigma}
\end{equation}
with the electron mass $m_e$ and the Boltzmann constant $k_B$.

Experimental uncertainties of the measured rate coefficients and cross sections typically amount to 10--15\% at a one-sigma confidence level. They stem predominantly from counting statistics and from systematic uncertainties of background subtraction, ion current and electron density determination.
As compared to single-pass experiments, the storage-ring technique offers the additional feature that pure ground-state beams of multiply charged ions can be prepared by storing the ions for a sufficiently long time before carrying out the ionization or recombination measurements. Thus, storage ring data are largely free from contributions by contaminating metastable beam components which frequently complicate the analysis of data from single-pass experiments.

\section{Electron-impact ionization of iron ions}

\begin{figure}[t]
\centering{\includegraphics[width=\textwidth]{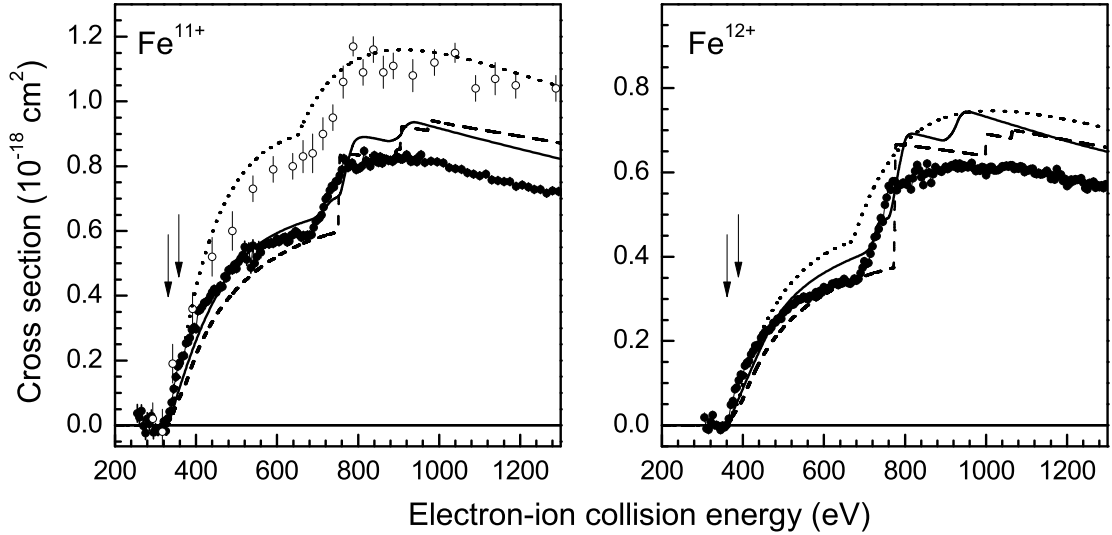}}%
\caption{\label{fig:FeEII}Cross sections for electron-impact ionization of Fe$^{11+}$ (left panel) and Fe$^{12+}$ ions (right panel): TSR experiments \protect\cite{Hahn2011,Hahn2011a} (closed symbols, uncertainties 12--16\% and 6--12\%, respectively), single-pass experiment  \protect\cite{Gregory1987a} (open symbols), theoretical calculations by Pindzola et al.\ \protect\cite{Pindzola1986a,Pindzola1987} (full lines) and by Dere \cite{Dere2007} (dashed lines), and  data from the compilation of Arnaud \& Raymond \protect\cite{Arnaud1992} (dotted lines). The  arrows mark the threshold energies for direct ionization by removal of a $3p$ or a $3s$ electron from the Fe$^{11+}$($3s^23p^3\;^4S_{3/2}$) ground-state (left panel) and the Fe$^{12+}$($3s^23p^2\;^3P_{0}$) ground-state (right panel). The steps in the cross sections are caused by excitation-autoionization processes which can occur only above certain threshold energies.}\end{figure}

The impact of this particular feature of the storage-ring technique can be seen in Figure \ref{fig:FeEII} (left panel) where the storage-ring EII cross section for Fe$^{11+}$ \cite{Hahn2011} is compared with the previous result from the single-pass experiment by Gregory et al.~\cite{Gregory1987a}. The cross section from this single-pass experiment is up to 40\% larger than the one from the storage-ring measurement. This deviation is larger than the combined experimental uncertainties of both measurements. It is most likely caused by the presence of unknown fractions of metastable ions in the ion beam of the single-pass experiment. The longest-lived metastable state of Fe$^{11+}$ is the $3s^23p^3\;^2D_{5/2}$ state with an excitation energy of 5.71~eV above the $3s^23p^3\;^4S_{3/2}$ ground state and a lifetime of about 0.5~s \cite{Ralchenko2010} which is much longer than the $\sim$10~$\mu$s flight time of the ions through the single-pass experiment.

In the storage-ring experiment, however, the ionization measurement started only after the ions had been stored for 2--3~s subsequent to their injection into the storage ring. By carefully modeling the populations of all metastable levels as function of storage time it was estimated that the sum of all populations of metastable levels amounted to less than 0.5\% when the ionization measurement was started. This value was even lower for the Fe$^{12+}$ EII measurement. The remaining uncertainties associated with such a low metastable ion-beam fraction do not contribute significantly to the error budget of the absolute cross section determination.

In comparison with results from distorted wave calculations \cite{Pindzola1986a,Pindzola1987,Dere2007} the single-pass Fe$^{11+}$ EII cross sections are significantly (up to about 40\%) larger than the theoretical results (Fig.~\ref{fig:FeEII}). This also holds for the data from the widely used compilation by Arnaud \& Raymond which are based on the results of the single-pass experiment. In contrast, the storage-ring cross-sections agree with the theoretical results within the experimental uncertainties at energies below the first excitation-autoionization thresholds. At higher energies the storage-ring cross sections are significantly smaller (by up to 18\%) than the theoretical values. These discrepancies which are somewhat larger than the experimental uncertainties (see caption of Fig.~\ref{fig:FeEII}) are most probably to be attributed to the theoretical uncertainties of the contributions by excitation-autoionization processes to the calculated total ionization cross sections. So far, the new-storage ring data lead to a more consistent picture of EII of iron ions than was available before.

\newpage

\begin{minipage}[htp]{\textwidth}
\centering{\includegraphics[width=0.8\textwidth]{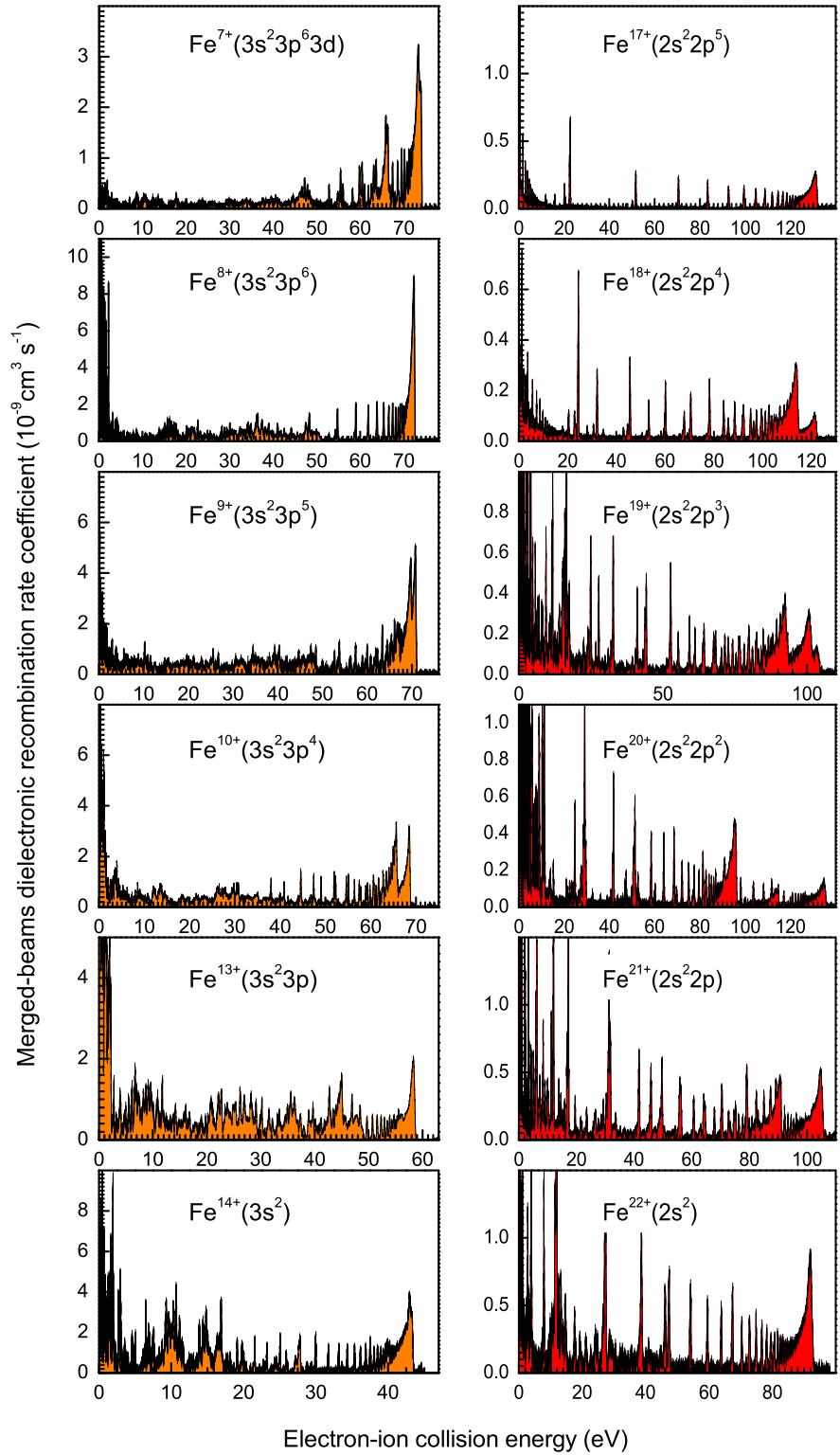}}%
\captionof{figure}{\label{fig:FeDR}(Color online) Measured merged-beams DR rate coefficients \protect\cite{SavinWiki} for some iron ions with open M-shell (left) and with open L-shell (right). For details see Refs.\ \protect\cite{Schmidt2008a} (Fe$^{7+}$, Fe$^{8+}$), \protect\cite{Lestinsky2009} (Fe$^{9+}$, Fe$^{10+}$), \protect\cite{Schmidt2006a} (Fe$^{13+}$), \protect\cite{Lukic2007a} (Fe$^{14+}$), \protect\cite{Savin1997,Savin1999}  (Fe$^{17+}$), \protect\cite{Savin1999,Savin2002c}  (Fe$^{18+}$), \protect\cite{Savin2002a}  (Fe$^{19+}$),  \protect\cite{Savin2003a} (Fe$^{20+}$, Fe$^{21+}$), \protect\cite{Savin2006a} (Fe$^{22+}$).}
\end{minipage}

\newpage

\section{Dielectronic recombination of iron ions}

Cosmic atomic plasmas can be divided into collisionally ionized plasmas (CP) and photoionized plasmas (PP) \cite{Savin2007d} both covering broad temperature ranges. Historically, most theoretical recombination data were calculated for CP \cite{Arnaud1992,Mazzotta1998} where highly charged ions exist only at rather large temperatures, e.g., in the solar corona. If the CP rate coefficients are also used for the astrophysical modeling of PP, inconsistencies arise. This has been noted, e.g., in the astrophysical modeling of X-ray spectra from active galactic nuclei (AGN) \cite{Netzer2004a,Kraemer2004a}.
It is clear, that these  deficiencies are due to a simplified theoretical treatment that was geared towards CP and more or less disregarded low-energy DR in order to keep the calculations tractable. Modern computers allow more sophisticated approaches, and recent theoretical work has aimed at providing a more reliable recombination data-base by  using state-of-the-art atomic codes \cite{Badnell2007b}.

\begin{figure}[b]
\centering{\includegraphics[width=0.6\textwidth]{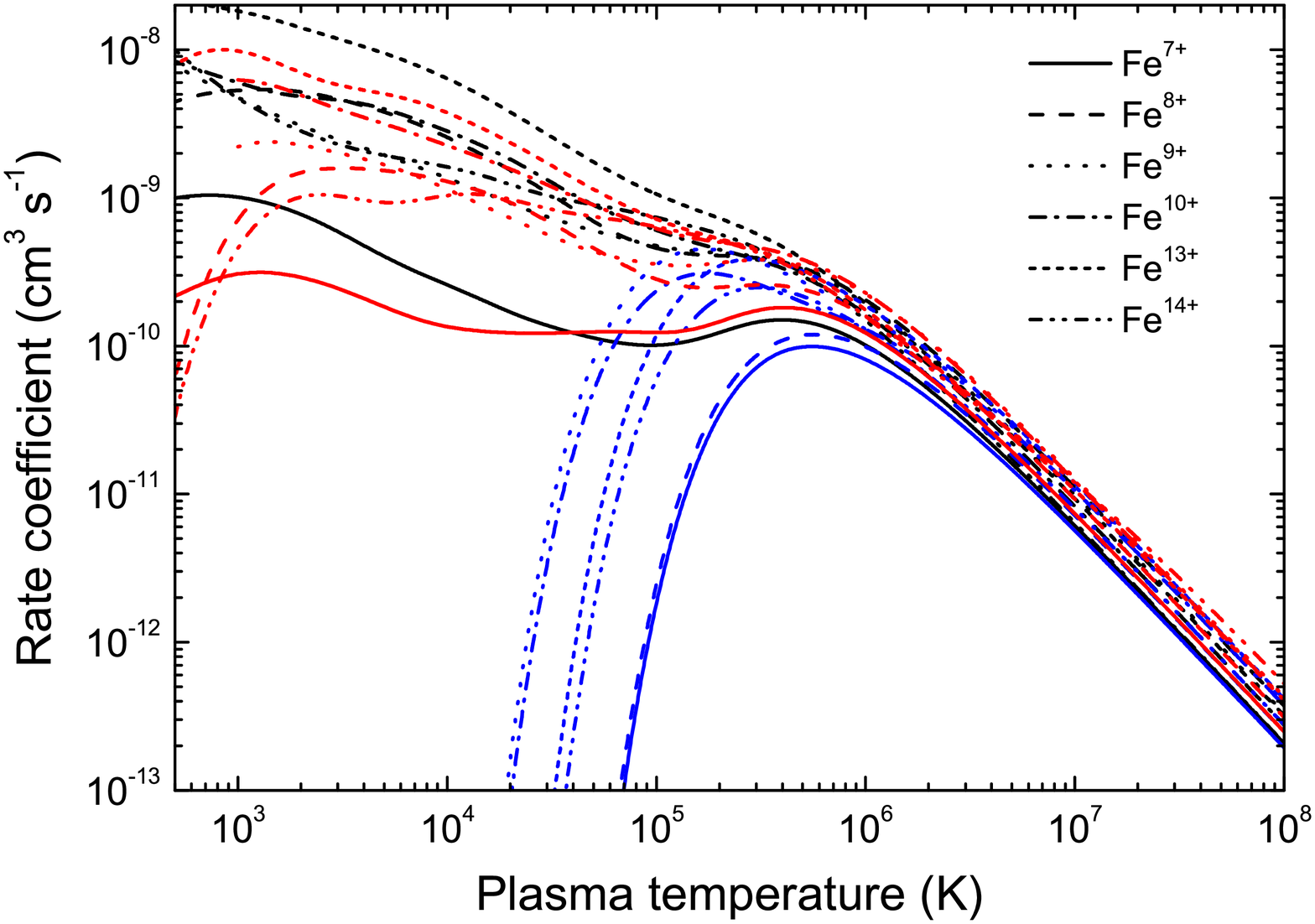}}%
\caption{\label{fig:FePlasma}(Color online) Plasma rate coefficients for the iron M-shell ions Fe$^{7+}$ (full lines), Fe$^{8+}$ (long-dashed lines), Fe$^{9+}$ (dotted lines), Fe$^{10+}$ (dash-dotted lines), Fe$^{13+}$ (short-dashed lines), and Fe$^{14+}$ (dash-dot-dotted lines). The experimentally derived curves \protect\cite{Schippers2010} are shown in black. The red curves are state-of-the-art theoretical results by Badnell \protect\cite{Badnell2006c} and the blue curves are from the compilation by Arnaud \& Raymond \protect\cite{Arnaud1992}}
\end{figure}

Corresponding experimental work on DR of iron ions has been carried out at the storage ring TSR over already more than one decade  (Fig.~\ref{fig:FeDR}). The experimental data resulting from this effort are presented in every detail in the references which are cited in the caption of Fig.~\ref{fig:FeDR}. A summary has been published recently \cite{Schippers2010}. Part of the resulting plasma DR rate coefficients, i.e., those for Fe M-shell ions which are available so far, are displayed in Fig.~\ref{fig:FePlasma} together with the corresponding widely-used rate coefficients from the compilation by Arnaud \& Raymond \cite{Arnaud1992} and with the corresponding state-of-the-art theoretical results by Badnell \cite{Badnell2006c}.

The discrepancies between the compiled rate coefficients on the one hand and the experimentally derived and state-of-the-art theoretical results on the other hand are striking. They are due the neglect of low-energy DR in the early theoretical work on which the compiled rate coefficients are based. As can be seen
from Fig.~\ref{fig:FeDR} most measured DR merged-beams rate coefficients are particularly strong at low electron-ion collision energies where, e.g., relativistic effects play a prominent role  (as discussion in more detail in \cite{Schippers2010}) which were entirely neglected in the early theoretical work. Although the results of the new theoretical work agrees much better with the experimental data than the early theoretical results, significant theoretical uncertainties remain (Fig.~\ref{fig:FePlasma}). Experimental benchmarks are thus indispensable for arriving at a reliable DR data base for the astrophysical modeling, particularly, of low-temperature plasmas.

Recent astrophysical model calculations \cite{Kallman2010} have made use of the new  DR rate coefficients. They reproduce the measured AGN x-ray spectrum much better than the above mentioned model calculations \cite{Netzer2004a,Kraemer2004a} which used the previously available rate coefficients from \cite{Arnaud1992}. One conclusion from the thus improved astrophysical model results is that the photoionized gas of an AGN is less highly charged than previously thought \cite{Kallman2010}.

\section{Dielectronic recombination of tungsten ions}

\begin{figure}[b]
\centering{\includegraphics[width=\textwidth]{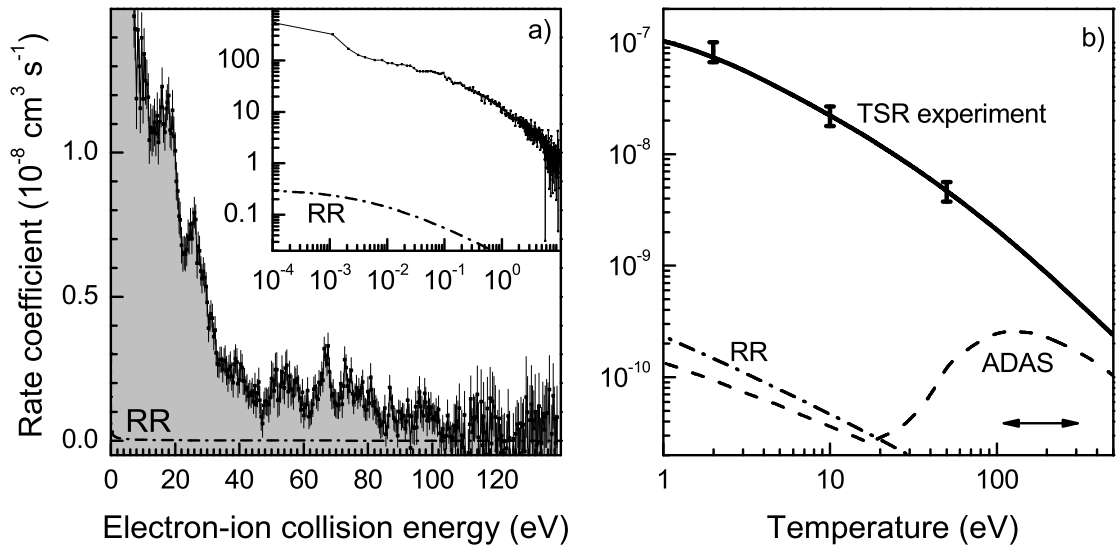}}%
\caption{\label{fig:W20}Dielectronic recombination of W$^{20+}$ \protect\cite{Schippers2011}: a) Measured merged-beams rate coefficient for electron-ion recombination of W$^{20+}$ as function of relative collision energy. The dash-dotted curve is the calculated RR rate coefficient using a hydrogenic approximation. The inset shows the same data in a log-log representation and a finer energy binning emphasizing the rate coefficient at very low energies. b) Rate coefficients for the recombination of W$^{20+}$ in a plasma. The thick full curve is the experimentally derived result comprising RR and DR with resonance energies below 140~eV. Error bars denote the 20\% experimental systematic uncertainty.  The dashed curve is the recombination rate coefficient from the ADAS data base  and the dash-dotted curve is again the result of the hydrogenic calculation for the RR rate coefficient. The horizontal arrow indicates the temperature range where W$^{20+}$ is expected to form in a fusion plasma.}
\end{figure}

Atomic spectroscopy and collision processes involving tungsten ions currently receive much attention, since tungsten is used as a wall material in nuclear fusion reactors \cite{Neu2009}. Consequently, tungsten ions are expected to be prominent impurities in fusion plasmas. Radiation from excited tungsten ions leads to substantial  plasma cooling which has to be well controlled in order to maintain the conditions for nuclear fusion.  Thus, a comprehensive knowledge of atomic energy levels and collision cross sections is required for a thorough understanding of the spatial and temporal evolution of the tungsten charge states and emission spectra in fusion plasmas \cite{Puetterich2010}. To date, only a small fraction of the needed atomic data has been derived from experimental measurements and most comes from theory \cite{Reader2009}. The situation is particularly problematic for electron-ion recombination since the DR rate coefficients from the ADAS data base \cite{adas} which are presently used for fusion plasma modeling are based on the semi-empirical Burgess formula \cite{Badnell2003a}. In order to improve this unsatisfying situation a dedicated research programme has been launched which aims at providing accurate experimental cross sections and rate coefficients for DR \cite{Schippers2011}, EII \cite{Rausch2011}, and PI \cite{Mueller2011} of tungsten ions.

Figure \ref{fig:W20}a displays our first storage-ring result, i.e., the merged-beams recombination rate coefficient for W$^{20+}$($4d^{10}4f^8$). Most dramatically, at energies at least up to 30~eV it is characterized by a high level about three orders of magnitude above the rate coefficient for non-resonant radiative recombination (RR) estimated from a hydrogenic calculation  \cite{Schippers2001c}. The measured rate coefficient decreases approximately monotonically from 0~eV up to an electron-ion collision energy of about 12 eV. From there on, broad resonance features become discernible up to the end of the experimental energy range. The widths of these features are much larger than the experimental energy spread. This indicates that the observed structures are most probably blends of many individually unresolved DR resonances. These findings are quite similar to what has been observed in a single-pass merged-beams experiment with isoelectronic Au$^{25+}$ ions \cite{Hoffknecht1998} where the measured rate coefficient also exceeded the RR rate coefficient by large factors. Because of the extraordinary complexity of the W$^{20+}$ atomic structure, no definitive assignment of the measured DR resonance features could  be made.

The huge DR resonances at very low electron-ion collision energies strongly influence the plasma rate coefficient even at temperatures above 100~eV as can been seen from Fig.~\ref{fig:W20}b. At all plasma temperatures the experimentally derived rate coefficient is very significantly larger than the W$^{20+}$ DR rate coefficient from the ADAS data base \cite{adas}. This difference amounts to a factor of 4.3 at a plasma temperature of 160~eV where the fractional abundance of W$^{20+}$ is predicted to peak in a fusion plasma. We expect similar discrepancies for other tungsten ions with open $4f$ shell and hope to stimulate rigorous theoretical investigations of DR for these ions.

\ack

The author would like to thank Alfred M\"{u}ller,  Daniel Savin, and Andreas Wolf for long-standing fruitful collaboration, and all his other collaborators who are too many to name them here (see references) for their contributions to this work. The excellent support by the MPIK accelerator and storage-ring crews is gratefully acknowledged. Financial support was provided in part by the Deutsche Forschungsgemeinschaft and by the Max-Planck-Society.
\medskip


\begin{thebibliography}{10}
\expandafter\ifx\csname url\endcsname\relax
  \def\url#1{{\tt #1}}\fi
\expandafter\ifx\csname urlprefix\endcsname\relax\def\urlprefix{URL }\fi
\providecommand{\eprint}[2][]{\url{#2}}

\bibitem{Ferland2003a}
Ferland G~J 2003 {\em Annu. Rev. Astron. Astrophys.\/} {\bf 41} 517--554

\bibitem{Kallman2007a}
Kallman T~R and Palmeri P 2007 {\em Rev. Mod. Phys.\/} {\bf 79} 79--133

\bibitem{Schippers2010}
Schippers S, Lestinsky M, M\"{u}ller A, Savin D~W, Schmidt E~W and Wolf A 2010
  {\em Int. Rev. At. Mol. Phys.\/} {\bf 1} 109--121 (\textit{Preprint}
  \eprint{arXiv:1002.3678v1})

\bibitem{Schippers2009a}
Schippers S 2009 {\em J. Phys.: Conf. Ser.\/} {\bf 163} 012001

\bibitem{Wolf2006c}
Wolf A, Buhr H, Grieser M, {von Hahn} R, Lestinsky M, Lindroth E, Orlov D~A,
  Schippers S and Schneider I~F 2006 {\em Hyperfine Interact.\/} {\bf 172}
  111--124

\bibitem{Mueller1999c}
M\"{u}ller A 1999 {\em Int. J. Mass Spectrom.\/} {\bf 192} 9--22

\bibitem{Schippers2001c}
Schippers S, M\"{u}ller A, Gwinner G, Linkemann J, Saghiri A~A and Wolf A 2001
  {\em Astrophys. J.\/} {\bf 555} 1027--1037

\bibitem{Schmidt2007b}
Schmidt E~W, Bernhardt D, M\"{u}ller A, Schippers S, Fritzsche S, Hoffmann J,
  Jaroshevich A~S, Krantz C, Lestinsky M, Orlov D~A, Wolf A, Luki\'c D and
  Savin D~W 2007 {\em Phys. Rev. A\/} {\bf 76} 032717

\bibitem{Lestinsky2009}
Lestinsky M, Badnell N~R, Bernhardt D, Grieser M, Hoffmann J, Luki\'c D,
  Schmidt E~W, M\"{u}ller A, Orlov D~A, Repnow R, Savin D~W, Schippers S, Wolf
  A and Yu D 2009 {\em Astrophys. J.\/} {\bf 698} 648--659

\bibitem{Hahn2010}
Hahn M, Bernhardt D, Lestinsky M, M\"{u}ller A, Schippers S, Wolf A and Savin
  D~W 2010 {\em Astrophys. J.\/} {\bf 712} 1166--1171

\bibitem{Schippers2000b}
Schippers S, Bartsch T, Brandau C, M\"{u}ller A, Gwinner G, Wissler G,
  Beutelspacher M, Grieser M, Wolf A and Phaneuf R~A 2000 {\em Phys. Rev. A\/}
  {\bf 62} 022708

\bibitem{Hahn2011}
Hahn M, Bernhardt D, Grieser M, Krantz C, Lestinsky M, M\"{u}ller A,
  Novotn\'{y} O, Repnow R, Schippers S, Wolf A and Savin D~W 2011 {\em
  Astrophys. J.\/} {\bf 729} 76

\bibitem{Hahn2011a}
Hahn M, Grieser M, Krantz C, Lestinsky M, M\"{u}ller A, Novotn\'{y} O, Repnow
  R, Schippers S, Wolf A and Savin D~W 2011 {\em Astrophys. J.\/} {\bf 735} 105

\bibitem{Gregory1987a}
Gregory, C D, Wang, J L, Meyer, W F, Rinn and K 1987 {\em Phys. Rev. A\/} {\bf
  35} 3256--3264

\bibitem{Pindzola1986a}
Pindzola M~S, Griffin D~C and Bottcher C 1986 {\em Phys. Rev. A\/} {\bf 34}
  3668 -- 3675

\bibitem{Pindzola1987}
Pindzola M~S, Griffin D~C, Bottcher C, Younger S~M and Hunter H~T 1987 {\em
  Nucl. Fusion Spec. Suppl.\/} {\bf 27} 21--41

\bibitem{Dere2007}
Dere K~P 2007 {\em Astron. Astrophys.\/} {\bf 466} 771--792

\bibitem{Arnaud1992}
Arnaud M and Raymond J 1992 {\em Astrophys. J.\/} {\bf 398} 394--406

\bibitem{Ralchenko2010}
Ralchenko Y, Kramida A, Reader J and {NIST ASD Team} 2010 Nist atomic spectra
  database (ver. 4.0), [online]. available:
  \url{http://www.nist.gov/physlab/data/asd.cfm} National Institute of
  Standards and Technology Gaithersburg, MD, USA
  
\bibitem{SavinWiki}
\url{https://docs.astro.columbia.edu/wiki/SavinGroup/DR}

\bibitem{Schmidt2008a}
Schmidt E~W, Schippers S, Bernhardt D, M\"{u}ller A, Hoffmann J, Lestinsky M,
  Orlov D~A, Wolf A, Luki\'c D~V, Savin D~W and Badnell N~R 2008 {\em Astron.
  Astrophys.\/} {\bf 492} 265--275

\bibitem{Schmidt2006a}
Schmidt E~W, Schippers S, M\"{u}ller A, Lestinsky M, Sprenger F, Grieser M,
  Repnow R, Wolf A, Brandau C, Luki\'c D, Schnell M and Savin D~W 2006 {\em
  Astrophys. J.\/} {\bf 641} L157--L160

\bibitem{Lukic2007a}
Luki\'c D, Savin D~W, Schnell M, Brandau C, Schmidt E~W, B\"{o}hm S, M\"{u}ller
  A, Schippers S, Lestinsky M, Sprenger F, Wolf A, Altun Z and Badnell N 2007
  {\em Astrophys. J.\/} {\bf 664} 1244--1252

\bibitem{Savin1997}
Savin D~W, Bartsch T, Chen M~H, Kahn S~M, Liedahl D~A, Linkemann J, M\"{u}ller
  A, Schippers S, Schmitt M, Schwalm D and Wolf A 1997 {\em Astrophys. J.\/}
  {\bf 489} L115--L118

\bibitem{Savin1999}
Savin D~W, Kahn S~M, Linkemann J, Saghiri A~A, Schmitt M, Grieser M, Repnow R,
  Schwalm D, Wolf A, Bartsch T, Brandau C, Hoffknecht A, M\"{u}ller A,
  Schippers S, Chen M~H and Badnell N~R 1999 {\em Astrophys. J. Suppl. Ser.\/}
  {\bf 123} 687--702

\bibitem{Savin2002c}
Savin D~W, Kahn S~M, Linkemann J, Saghiri A~A, Schmitt M, Grieser M, Repnow R,
  Schwalm D, Wolf A, Bartsch T, M\"{u}ller A, Schippers S, Chen M~H, Badnell
  N~R, Gorczyca T~W and Zatsarinny O 2002 {\em Astrophys. J.\/} {\bf 576}
  1098--1107

\bibitem{Savin2002a}
Savin D~W, Behar E, Kahn S~M, Gwinner G, Saghiri A~A, Schmitt M, Grieser M,
  RRepnow, Schwalm D, Wolf A, Bartsch T, M\"{u}ller A, Schippers S, Badnell
  N~R, Chen M~H and Gorczyca T~W 2002 {\em Astrophys. J. Suppl. Ser.\/} {\bf
  138} 337--370

\bibitem{Savin2003a}
Savin D~W, Kahn S~M, Gwinner G, Grieser M, Repnow R, Saathoff G, Schwalm D,
  Wolf A, M\"{u}ller A, Schippers S, Z\'{a}vodsky P~A, Chen M~H, Gorczyca T~W,
  Zatsarinny O and Gu M~F 2003 {\em Astrophys. J. Suppl. Ser.\/} {\bf 147}
  421--435

\bibitem{Savin2006a}
Savin D~W, Gwinner G, Grieser M, Repnow R, Schnell M, Schwalm D, Wolf A, Zhou
  S~G, Kieslich S, M\"{u}ller A, Schippers S, Colgan J, Loch S~D, Chen M~H and
  Gu M~F 2006 {\em Astrophys. J.\/} {\bf 642} 1275--1285

\bibitem{Savin2007d}
Savin D~W 2007 {\em J. Phys.: Conf. Ser.\/} {\bf 88} 012071

\bibitem{Mazzotta1998}
Mazzotta P, Mazzitelli G, Colafrancesco S and Vittorio N 1998 {\em Astron.
  Astrophys.\/} {\bf 133} 403--409

\bibitem{Netzer2004a}
Netzer H 2004 {\em Astrophys. J.\/} {\bf 604} 551--555

\bibitem{Kraemer2004a}
Kraemer S~B, Ferland G~J and Gabel J~R 2004 {\em Astrophys. J.\/} {\bf 604}
  556--561

\bibitem{Badnell2007b}
Badnell N~R 2007 {\em J. Phys.: Conf. Ser.\/} {\bf 88} 012070

\bibitem{Badnell2006c}
Badnell N~R 2006 {\em Astrophys. J.\/} {\bf 651} L73--L76

\bibitem{Kallman2010}
Kallman T 2010 {\em Space Science Reviews\/} {\bf 157} 177--191

\bibitem{Schippers2011}
Schippers S, Bernhardt D, M\"{u}ller A, Krantz C, Grieser M, Repnow R, Wolf A,
  Lestinsky M, Hahn M, Novotn\'{y} O and Savin D~W 2011 {\em Phys. Rev. A\/}
  {\bf 83} 012711

\bibitem{Neu2009}
Neu R, Bobkov V, Dux R, Fuchs J~C, Gruber O, Herrmann A, Kallenbach A, Maier H,
  Mayer M, P\"{u}tterich T, Rohde V, Sips A~C~C, Stober J, Sugiyama K and
  {ASDEX Upgrade Team} 2009 {\em Phys. Scr.\/} {\bf T138} 014038

\bibitem{Puetterich2010}
P\"{u}tterich T, Neu R, Dux R, Whiteford A, O'Mullane M, Summers H and the
  ASDEX Upgrade~Team 2010 {\em Nucl. Fusion\/} {\bf 50} 025012

\bibitem{Reader2009}
Reader J 2009 {\em Phys. Scr.\/} {\bf T134} 014023

\bibitem{adas}
\url{http://www.adas.ac.uk/}

\bibitem{Badnell2003a}
Badnell N~R, O'Mullane M~G, Summers H~P, Altun Z, Bautista M~A, Colgan J,
  Gorczyca T~W, Mitnik D~M, Pindzola M~S and Zatsarinny O 2003 {\em Astron.
  Astrophys.\/} {\bf 406} 1151--1161

\bibitem{Rausch2011}
Rausch J, Becker A, Spruck K, Hellhund J, {Borovik Jr} A, Huber K, Schippers S
  and M\"{u}ller A 2011 {\em J. Phys. B\/} {\bf 44} 165202

\bibitem{Mueller2011}
M\"{u}ller A, Schippers S, Kilcoyne A~L~D and Esteves D 2011 {\em Phys. Scr.\/}
  {\bf T144} 014052

\bibitem{Hoffknecht1998}
Hoffknecht A, Uwira O, Frank A, Schennach S, Spies W, Wagner M, Schippers S,
  M\"{u}ller A, Becker R, Kleinod M, Angert N and Mokler P~H 1998 {\em J. Phys.
  B\/} {\bf 31} 2415--2428

\end{thebibliography}
\providecommand{\newblock}{}

\end{document}